\documentclass[article, twocolumn]{revtex4}
\usepackage{graphicx,amsmath, amssymb, bm, epstopdf,dsfont}

\begin{document}

\title{Single-shot measurement of the orbital-angular-momentum spectrum of light}

\author{Girish Kulkarni$^{1}$, Rishabh Sahu$^{1}$, Omar S. Maga$\tilde{\mathrm{n}}$a-Loaiza$^{2}$, Robert. W. Boyd$^{2,3}$, and Anand K. Jha$^{1}$}

\email{akjha9@gmail.com}

\affiliation{$^{1}$Department of Physics, Indian Institute of Technology, Kanpur, 208016, India\\
$^{2}$The Institute of Optics, University of Rochester, Rochester, New York 14627, USA\\
$^{3}$Department of Physics, University of Ottawa, Ottawa, ON, K1N6N5, Canada}

\date{\today}

\begin{abstract}

The existing methods for measuring the orbital-angular-momentum (OAM) spectrum suffer from issues such as poor efficiency, strict
interferometric stability requirements, and too much loss. Furthermore, most techniques inevitably discard
part of the field and measure only a post-selected portion of the
true spectrum. Here, we propose and demonstrate a new interferometric technique for measuring the true OAM spectrum of optical fields in a single-shot manner.
Our technique directly encodes the OAM-spectrum information in the
azimuthal intensity profile of the output interferogram. In
the absence of noise, the spectrum can be fully decoded using a single
acquisition of the output interferogram, and, in the presence
of noise, acquisition of two suitable interferograms is sufficient for the purpose. As an important
application of our technique, we demonstrate measurements of the
angular Schmidt spectrum of the entangled photons produced by
parametric down-conversion and report a broad spectrum with the angular Schmidt number 82.1.

\end{abstract}

\maketitle

It was shown by Allen {\it et al.} that a photon in a light beam
can have orbital angular momentum (OAM) values in the integer
multiples of $\hbar$ \cite{allen1992pra}. This result has made OAM
a very important degree of freedom for both classical and quantum
information protocols \cite{bennett1984, bennett92prl, ekert91prl,
duan2001nature, marcikic2003long, ladd2010quantum,
kimble2008quantum,
wang2012natphot,bozinovic2013science,yan2014natcom}. This is
because an information protocol requires a discrete basis, and the
OAM degree of freedom provides a basis that is not only discrete
but can also be high-dimensional \cite{barnett1990pra,
yao2006opex, jha2008pra, jack2008njp}. This is in contrast to the
polarization degree of freedom, which provides a discrete but only
a two-dimensional basis \cite{bennett1984, bennett92prl,
ekert91prl}. High-dimensional quantum information protocols have
many distinct advantages in terms of security
\cite{karimipour2002pra,cerf2002prl,nikolopoulos2006pra},
transmission bandwidth \cite{fujiwara2003prl,cortese2004pra}, gate
implementations \cite{ralph2007pra,lanyon2009natphy},
supersensitive measurements \cite{jha2011pra2} and fundamental
tests of quantum mechanics
\cite{kaszlikowski2000prl,collins2002prl,vetesi2010prl,
leach2009optexp}. In the classical domain, the high-dimensional
OAM-states can increase the system capacities and spectral
efficiencies
\cite{wang2012natphot,bozinovic2013science,yan2014natcom}.

One of the major challenges faced in the implementation of
OAM-based high-dimensional protocols is the efficient detection of
OAM spectrum, and it is currently an active research area
\cite{mair2001nature, heckenberg1992optlett, zambrini2006prl,
pires2010opl, pires2010prl,jha2011pra,malik2012pra,
vasnetsov2003optlett,zhou2017lsa, leach2002prl}. There are several
approaches to measuring the OAM spectrum of a field. One main
approach \cite{mair2001nature, heckenberg1992optlett} is to
display a specific hologram onto a spatial light modulator (SLM)
for a given input OAM-mode and then measure the intensity at the
first diffraction order using a single mode fiber. This way, by
placing different holograms specific to different input OAM-mode
in a sequential manner, one is able to measure the spectrum.
However, this method is very inefficient since the required number
of measurements scales with the size of the input spectrum.
Moreover, due to the non-uniform fiber-coupling efficiencies of
different input OAM-modes \cite{qassim2014josab}, this method does
not measure the true OAM spectrum. The second approach relies on
measuring the angular coherence function of the field and then
reconstructing the OAM-spectrum through an inverse Fourier
transform. One way to measure the angular coherence function is by
measuring the interference visibility in a Mach-Zehnder
interferometer as a function of the Dove-prism rotation angle
\cite{pires2010opl, pires2010prl}. Although this method does not
have any coupling-efficiency issue, it still requires a series of
measurements for obtaining the angular coherence function. This
necessarily requires that the interferometer be kept aligned for
the entire range of the rotation angles. A way to bypass the
interferometric stability requirement is by measuring the angular
coherence function \cite{jha2011pra, malik2012pra} using angular
double-slits \cite{jha2010prl}. However, this method also requires
a series of measurements and since in this method only a very
small portion of the incident field is used for detection, it is
not suitable for very low-intensity fields such as the fields
produced by parametric down-conversion (PDC). The other approaches
to measuring the OAM spectrum include techniques based on
rotational Doppler frequency shift \cite{vasnetsov2003optlett,
zhou2017lsa} and concatenated Mach-Zehnder interferometers
\cite{leach2002prl}. However, due to several experimental
challenges, these approaches \cite{vasnetsov2003optlett,
zhou2017lsa, leach2002prl} have so far been demonstrated only for
fields consisting of just a few modes. Thus the existing methods for measuring the OAM spectrum
information suffer from either poor efficiency
\cite{mair2001nature, vasnetsov2003optlett} or strict
interferometric stability requirements \cite{pires2010opl,
pires2010prl, leach2002prl} or too much loss \cite{jha2011pra,
malik2012pra}. In addition, many techniques
\cite{mair2001nature,jha2011pra, malik2012pra} inevitably discard
part of the field and yield only a post-selected portion of the
true spectrum. 

In this article, we demonstrate a novel
interferometric technique for measuring the true OAM spectrum in a
single-shot manner, that is, by acquiring only one image of the output interferogram using a multi-pixel camera. Since our method
is interferometric, the efficiency is very high, and since it
involves only single-shot measurements, the interferometric
stability requirements are much less stringent.

\

\noindent {\bf Results}

\noindent {\bf Theory of single-shot spectrum measurement.} The Laguerre-Gaussian (LG) modes, represented as
$LG_p^l(\rho, \phi)$, are exact solutions of the paraxial
Helmholtz equation. The OAM-mode index $l$ measures the OAM of
each photon in the units of $\hbar$, while the index $p$
characterizes the radial variation in the intensity
\cite{allen1992pra}. The partially coherent fields that we
consider in this article are the ones that can be represented as
incoherent mixtures of LG modes having different OAM-mode indices
\cite{jha2011pra}. The electric field $E_{\rm in}(\rho, \phi)$
corresponding to such a field can be written as
\begin{align}
E_{\rm in}(\rho, \phi)=\sum_{l, p} A_{lp} LG_p^l(\rho,
\phi)=\sum_{l, p} A_{lp} LG_p^l(\rho)e^{il\phi}, \label{field}
\end{align}
where $A_{lp}$ are stochastic variables. The corresponding
correlation function $W(\rho_1,\phi_1; \rho_2, \phi_2)$ is
\begin{align}
W(\rho_1,\phi_1; \rho_2, &\phi_2)\equiv \langle E_{\rm
in}^*(\rho_1, \phi_1)E_{\rm in}(\rho_2,
\phi_2)\rangle_e \notag\\
&=\sum_{l, p, p'} \alpha_{l p p'} LG^{*l}_p(\rho_1,
\phi_1)LG^{l}_{p'}(\rho_2, \phi_2). \label{coherence-fn}
\end{align}
Here $\langle\cdots\rangle_e$ represents the ensemble average and
$\langle A^*_{lp}A_{l'p'} \rangle_e=\alpha_{l p p'}\delta_{l,l'}$,
where $\delta_{l,l'}$ is the Kronecker-delta function. When
integrated over the radial coordinate, the above correlation
function yields the angular coherence function: $W(\phi_1,
\phi_2)\equiv \int_0^{\infty} \rho d\rho W(\rho,\phi_1; \rho,
\phi_2)$, which, for the above field, can be shown to be
\cite{jha2011pra}
\begin{align}
W(\phi_1, \phi_2)\rightarrow W(\Delta\phi)
=\frac{1}{2\pi}\sum_{l=-\infty}^{\infty}S_{l} e^{-il\Delta\phi},
\label{angular-coherence-fn}
\end{align}
where $S_l=\sum_{p}\alpha_{l p p}$, $\Delta\phi=\phi_1-\phi_2$,
and where we have used the identity $\int_0^{\infty}\rho
LG_{p}^{*l}(\rho)LG_{p'}^{l}(\rho){\rm d}\rho=\delta_{pp'}/2\pi $.
The quantity $S_l$ is referred to as the OAM spectrum of the
field. It is normalized such that $\sum_l S_l=1$ and
$\int_{-\pi}^{\pi} W(\phi_1, \phi_1) d\phi_1=\int_{-\pi}^{\pi}
W(\phi_2, \phi_2) d\phi_2=1$. The Fourier transform relation of
Equation (\ref{angular-coherence-fn}) is the angular analog of the
temporal Wiener-Khintchine theorem for temporally-stationary
fields (see Section 2.4 of \cite{mandel1995optical}). Therefore, a
measurement of $W(\Delta\phi)$ can yield the OAM spectrum of the
input field through the inverse Fourier relation
\begin{align}
S_l  =\int_{-\pi} ^{\pi} W(\Delta\phi) e^{il\Delta\phi}
d\Delta\phi. \label{OAM-spectrum}
\end{align}
\begin{figure*}[t!]
\hspace*{2.5mm}
\includegraphics[width=179mm,keepaspectratio]{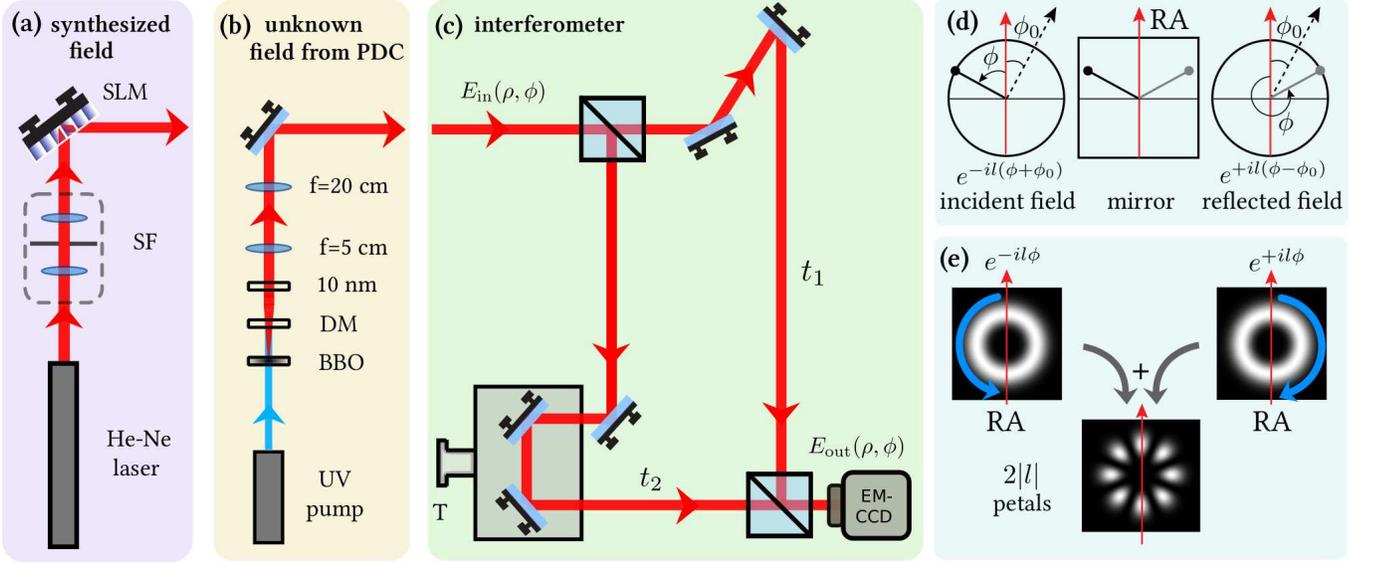}
\vspace{-4mm} \caption{{\bf Describing the proposed experimental technique and its working priciple.} ({\bf a}) Schematic of the setup for synthesizing
 partially coherent field of the
type represented by Equations (\ref{field}) and (\ref{coherence-fn})
with known OAM spectra. ({\bf b}) Schematic of the setup for producing
the partially coherent fields through parametric down-conversion.
({\bf c}) Schematic of the Mach-Zehnder interferometer. ({\bf d}) Describing
how a mirror reflection changes the azimuthal phase of an LG mode.
An incident beam with the azimuthal phase profile
$e^{il(\phi+\phi_0)}$ transforms into a beam having the azimuthal
phase profile $e^{-il(\phi-\phi_0)}$, where $\phi_0$ is the angle
between the reflection axis (RA) and the zero-phase axis (dashed
axis) of the incident mode. ({\bf e}) Illustrating the interference
effect produced by the interferometer when the incident field is
an $LG_{p=0}^{l}(\rho, \phi)$ mode with $l=4$. At the output, we
effectively have the interference of an $e^{il\phi}$ mode with an
$e^{-il\phi}$ mode, and we obtain the output interference
intensity in the form of a petal pattern with the number of petals
being $2|l|=8$. SLM: spatial light modulator; SF: spatial filter;
DM: dichroic mirror; BBO: type-I beta barium borate crystal; T:
translation stage.}\label{setup-schematic}
  \vspace{-1mm}
\end{figure*}
Now, let us consider the situation shown in
Fig.~\ref{setup-schematic}(a). A partially coherent field of the
type represented by Equations (\ref{field}) and (\ref{coherence-fn})
enters the Mach-Zehnder interferometer having an odd and an even
number of mirrors in the two arms [shown in
Fig.~\ref{setup-schematic}(c) ]. As illustrated in
Fig.~\ref{setup-schematic}(d), each reflection transforms the
polar coordinate as $\rho \to \rho$ and the azimuthal coordinate
as $\phi+\phi_0 \to -\phi+\phi_0$ across the reflection axis (RA).
Here $\phi$ is the angle measured from RA, and $\phi_0$ is the
angular-separation between RA and the zero-phase axis of the
incident mode (dashed-axis). The phase $\phi_0$ does not survive
in intensity expressions. So, without the loss of any generality,
we take $\phi_0=0$ for all incident modes. Therefore, for the
input incident field $E_{\rm in}(\rho, \phi)$ of
Equation (\ref{field}), the field $E_{\rm out}(\rho, \phi)$ at the
output port becomes
\begin{align}
E_{\rm out}(\rho, \phi)=\sqrt{k_1} &E_{\rm in}(\rho,
-\phi)e^{i(\omega_0 t_1+\beta_1)}\notag\\
+&\sqrt{k_2} E_{\rm in}(\rho, \phi)e^{i(\omega_0
t_2+\beta_2+\tilde\gamma)}.
\end{align}
Here, $t_{1}$ and $t_{2}$ denote the travel-times in the two arms
of the interferometer; $\omega_0$ is the central frequency of the
field; $\beta_{1}$ and $\beta_{2}$ are the phases other than the
dynamical phase acquired in the two arms; $\tilde{\gamma}$ is a
stochastic phase which incorporates the temporal coherence between
the two arms; $k_1$ and $k_2$ are the scaling constants in the two
arms, which depend on the splitting ratios of the beam splitters,
etc. The azimuthal intensity $I_{\rm out}(\phi)$ at the output
port is defined as $I_{\rm out}(\phi)\equiv\int \rho  \langle
E_{\rm out}^*(\rho, \phi) E_{\rm out}^*(\rho, \phi) \rangle_e
d\rho$, and using Equations (\ref{field})-(\ref{OAM-spectrum}), we can
evaluate it to be
\begin{align}
I_{\rm out}(\phi)=\frac{1}{2\pi}(k_1+k_2)+\gamma \sqrt{k_1k_2}
W(2\phi)e^{i\delta}+{\rm c.c.} \label{output-intensity}
\end{align}
Here, we have defined $\delta\equiv
\omega_0(t_{2}-t_{1})+(\beta_{2}-\beta_{1})$, and $\gamma=\langle
e^{i\tilde{\gamma}}\rangle$ quantifies the degree of temporal
coherence. The intensity expression in
Equation (\ref{output-intensity}) is very different from the output
intensity expression one obtains in a conventional Mach-Zehnder
interferometer with a Dove prism having either odd/odd or
even/even number of mirrors in the two interferometric arms
\cite{pires2010opl,pires2010prl}. In Equation (\ref{output-intensity}),
the output intensity and the angular correlation function both
depend on the detection-plane azimuthal angle $\phi$. As a result,
the angular correlation function $W(2\phi)$ comes out encoded in
the azimuthal intensity profile $I_{\rm out}(\phi)$. In contrast, in
the conventional Mach-Zehnder interferometers
\cite{pires2010opl,pires2010prl}, the output intensity has no
azimuthal variation; one measures the angular correlation function
by measuring the interference-visibility of the total output
intensity as a function of the Dove prism rotation angles. For a
symmetric spectrum $\left[S_l=S_{-l}=(S_l+S_{-l})/2\right]$,  we
have using the formula in Equation \ref{OAM-spectrum}
\begin{align}
\!\! S_l&=\int_{-\pi}^{\pi} W(2\phi)e^{i2l\phi}d(2\phi) \notag\\
&=\frac{1}{2\gamma\cos\delta\sqrt{k_1k_2}}
\int_{-\pi}^{\pi}\left[I_{\rm
out}(\phi)-\frac{k_1+k_2}{2\pi}\right]
\cos{(2l\phi)}d\phi.\label{single-shot spectrum}
\end{align}
So, if the precise values of $k_1,k_2,\gamma$ and $\delta$ are
known then a single-shot measurement of the output interferogram
$I_{\rm out}(\phi)$ yields the angular coherence function
$W(2\phi)$ and thereby the OAM spectrum $S_l$. Here, by ``a single-shot measurement" we mean recording one image of the output interferogram using a multi-pixel camera. The recording may involve collecting several photons per pixel for a fixed exposure time of the camera.

\begin{figure*}[t!]
 \hspace*{-1.5mm}
\includegraphics[width=173 mm,keepaspectratio]{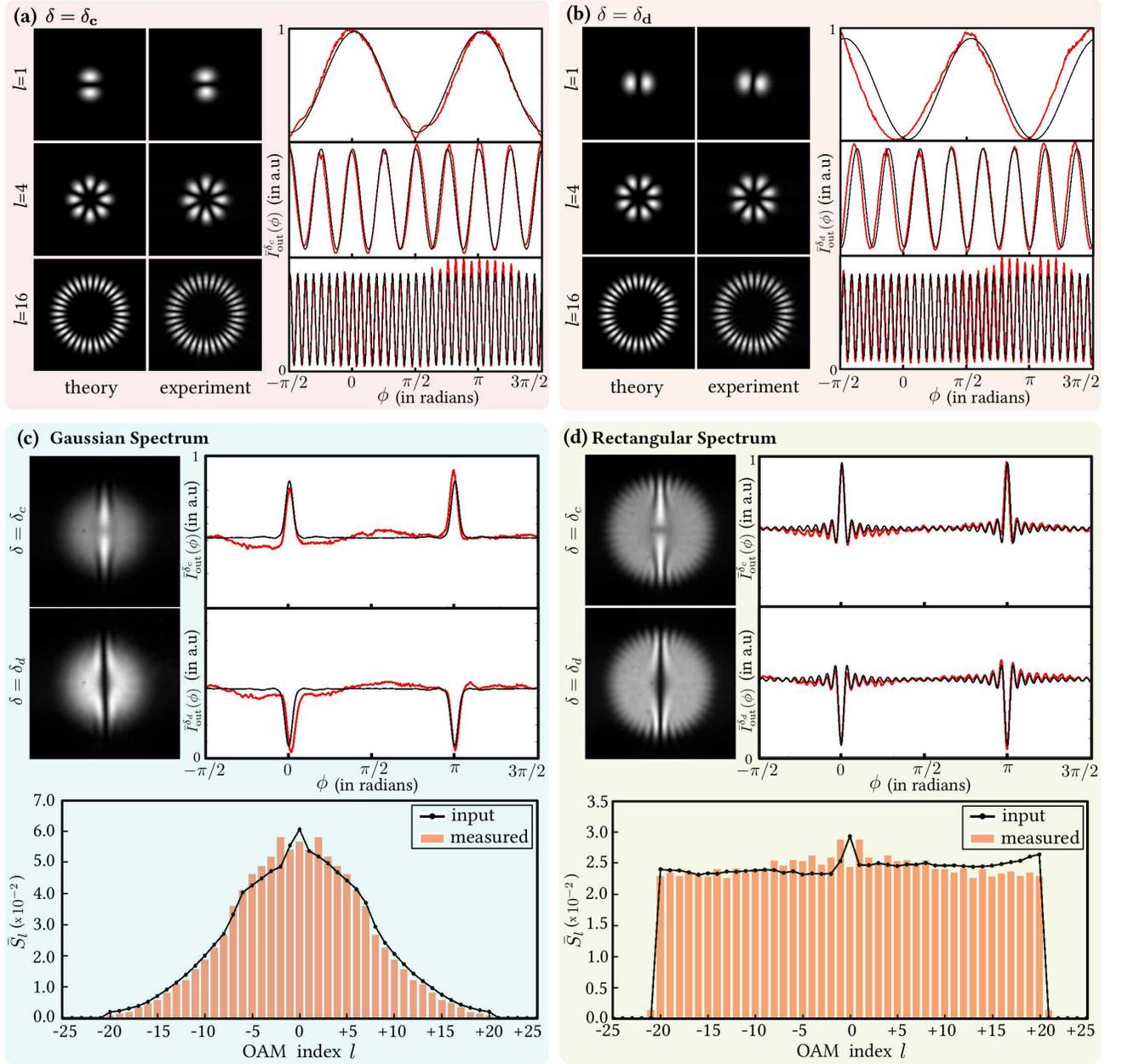}
 \vspace*{-3mm}
\caption{{\bf Experimental results obtained with the lab-synthesized fields.} ({\bf a}) and ({\bf b}) Measured output interferograms and the
corresponding azimuthal intensities for input $LG_{p=0}^l(\rho,
\phi)$ modes with $l=1,4,$ and $16$ for $\delta_{c}\approx 2m\pi$
and $\delta_{d}\approx(2m+1)\pi$, respectively, where $m$ is an
integer.  ({\bf c}) Measured output interferograms, the azimuthal
intensities, and the measured spectrum for the synthesized input
field with a Gaussian OAM-spectrum. ({\bf d}) Measured output
interferograms, the azimuthal intensities, and the measured
spectrum for the synthesized input field with a Rectangular
OAM-spectrum. In all the above azimuthal intensity plots, the red lines are the experimental plots and the black lines are the theoretical fits.} \label{observations} \vspace*{-1mm}
\end{figure*}

\

\noindent {\bf Theory of two-shot noise-insensitive spectrum
measurement.} Although it is in principle possible to measure the OAM
spectrum in a single-shot manner as discussed above, it is
practically extremely difficult to do so because of the
requirement of a very precise knowledge of $k_1,k_2,\gamma$, and
$\delta$. Moreover, obtaining a spectrum in this manner is
susceptible to noise in the measured $I_{\rm out}(\phi)$, which
results in errors in the measured spectrum. We now show that it is
possible to eliminate this noise completely while also
relinquishing the need for a precise knowledge of $k_1$, $k_2,
\gamma$, and $\delta$, just by acquiring one additional output
interferogram. We present our analysis for a symmetric spectrum,
that is, for $\left[S_l=S_{-l}=(S_l+S_{-l})/2\right]$. (See
Methods section for the non-symmetric case). Let us assume
that the experimentally measured output azimuthal intensity
$\bar{I}_{\rm out}^\delta(\phi)$ at $\delta$ contains some noise
$I_{\rm n}^{\delta}(\phi)$ in addition to the signal $I_{\rm
out}(\phi)$, that is,
\begin{align}
\bar{I}_{\rm out}^{\delta}(\phi)=I_{\rm n}^{\delta}(\phi)
+\frac{1}{2\pi}(k_1+k_2)+2\gamma \sqrt{k_1k_2}
W(2\phi)\cos\delta.\notag
\end{align}
Now, suppose that we have two interferograms, $\bar{I}_{\rm out}^{
\delta_c}(\phi)$ and $\bar{I}_{\rm out}^{ \delta_d}(\phi)$,
measured at $\delta=\delta_c$ and $\delta=\delta_d$, respectively.
The difference in the intensities $\Delta \bar{I}_{\rm
out}(\phi)=\bar{I}_{\rm out}^{ \delta_c}(\phi)-\bar{I}_{\rm out}^{
\delta_d}(\phi)$ of the two interferograms is then given by
\begin{align}
\!\!\!\Delta \bar{I}_{\rm out}(\phi)=\Delta I_{\rm
n}(\phi)+2\gamma \sqrt{k_1k_2} (\cos\delta_c -
\cos\delta_d)W(2\phi),\notag
\end{align}
where $\Delta I_{\rm n}(\phi)=I_{\rm n}^{\delta_c}(\phi) - I_{\rm
n}^{\delta_d}(\phi)$ is the difference in the noise intensities.
Multiplying each side of the above equation by $e^{i2l\phi}$,
using the formula in Equation \ref{OAM-spectrum}, and defining the
measured OAM spectrum as $\bar{S_l}\equiv \int_{-\pi}^{\pi}\Delta
\bar{I}_{\rm out}(\phi) e^{i2l\phi}
d(2\phi)=\int_{-\pi}^{\pi}\Delta \bar{I}_{\rm out}(\phi)
e^{i2l\phi} d\phi$, we get
\begin{align}
\!\!\!\!\bar{S_l}=\int_{-\pi}^{\pi}\!\!\!\!\Delta {I}_{\rm
n}(\phi) e^{i2l\phi} d\phi+2\gamma \sqrt{k_1k_2} (\cos\delta_c -
\cos\delta_d)S_l. \label{measured-spectrum1}
\end{align}
We see that in situations in which there is no shot-to-shot
variation in noise, that is, $\Delta I_{\rm n}(\phi)=0$, the
measured OAM-spectrum $\bar{S_l}$ is same as the true input
OAM-spectrum $S_l$ up to a scaling constant. One can thus obtain
the normalized OAM-spectrum in a two-shot manner without having to
know the exact values of $k_1$, $k_2, \gamma, \delta_c$ or
$\delta_d$. Nevertheless, in order to get a better experimental
signal-to-noise ratio, it would be desirable to have
$\gamma\approx 1$, $k_1\approx k_2\approx 0.5$, $\delta_c\approx
0$, and $\delta_d\approx \pi$. Now, in situation in which $\Delta
I_{\rm n}(\phi)\neq0$, it is clear from
Equation (\ref{measured-spectrum1}) that the measured spectrum will
have extra contributions. However, since we do not expect very
rapid azimuthal variations in $\Delta I_{\rm n}(\phi)$, the extra
contributions should be more prominent for modes around $l=0$ and should
die down for large-$l$ modes.
\begin{figure*}[t!]
 \hspace*{-1.5mm}
\includegraphics[width=182mm,keepaspectratio]{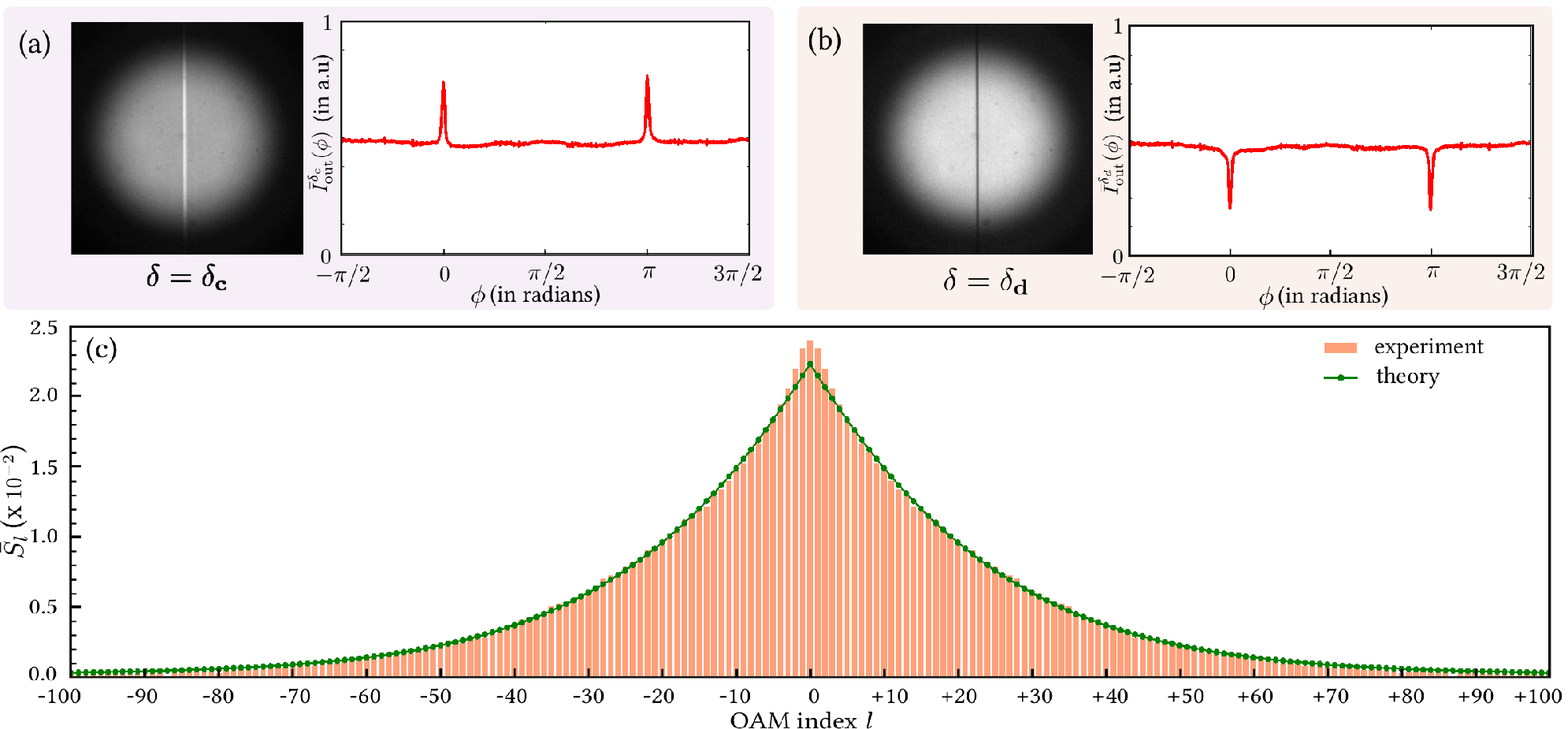}
  \vspace*{-6.5mm}
\caption{{\bf Experimental results obtained with fields produced by parametric down-conversion.} ({\bf a}) and ({\bf b}) Measured output interferogram and the
azimuthal intensity for $\delta=\delta_{c}$ and
$\delta=\delta_{d}$, respectively. ({\bf c}) The normalized measured
spectrum $\bar{S}_{l}$ as computed using
Equation (\ref{measured-spectrum1}), and the normalized theoretical
spectrum as calculated using the formalism of
Ref.~\cite{torres2003pra}, for our setup parameters, namely, a
type-I collinear down-conversion with a 2-mm thick BBO crystal and
a $0.85$-mm beam-waist pump laser. The theoretical spectrum has no
fitting parameters. The angular Schmidt number
$K=1/(\sum_{l}\bar{S}^{2}_{l})$ for the measured spectrum is
evaluated to be $82.1$.  } \label{spdcpatterns} \vspace*{-2mm}
\end{figure*}

\

\noindent {\bf Measuring lab-synthesized OAM spectra.} We now report the experimental demonstrations of our
technique for laboratory-synthesized, symmetric OAM spectra. As
shown in Fig.~\ref{setup-schematic}(a), a He-Ne laser is spatially
filtered and made incident onto a Holoeye Pluto SLM. The
$LG_{p=0}^l(\rho, \phi)$ modes are generated using the method by
Arrizon {\it et al.} \cite{arrizon2007josaa} and then made
incident into the Mach-Zehnder interferometer shown in
Fig.~\ref{setup-schematic}(c). The measured interferogram and the
corresponding azimuthal intensity $\bar{I}_{\rm
out}^{\delta}(\phi)$ for a few $LG_{p=0}^l(\rho, \phi)$ modes for
$\delta_c\approx2m\pi$, and $\delta_d\approx(2m+1)\pi$, where $m$
is an integer, are presented in Fig.~\ref{observations}(a) and
Fig.~\ref{observations}(b), respectively. A very good match
between the theory and experiment indicates that the
$LG_{p=0}^l(\rho, \phi)$ modes produced in our experiments are of
very high quality. By controlling the strengths of the synthesized
$LG_{p=0}^l(\rho, \phi)$ modes for $l$ ranging from $l=-20$ to
$l=20$, we synthesize two separate fields: one with a rectangular
spectrum and the other one with a Gaussian spectrum. The
representative interferogram corresponding to a particular field
as input is obtained by adding individual interferograms for $l$
ranging from $l=-20$ to $l=+20$. Two such representative
interferograms, one with $\delta=\delta_c$ and other one with
$\delta=\delta_d$ are recorded for each field.
Figures \ref{observations}(c) and \ref{observations}(d) show the
measured output interferograms, the corresponding azimuthal
intensities, and the measured spectrum $\bar{S}_l$ computed using
Equation (\ref{measured-spectrum1}) for the synthesized Gaussian and
Rectangular OAM spectrum, respectively. We find a very good match
between the synthesized spectra and the measured spectra. There is
some mismatch in the measured spectra for low-$l$ modes. We
attribute this to SLM imperfections, various wave-front
aberrations, and the non-zero shot-to-shot noise variation $\Delta
I_{\rm n}(\phi)$.

\

\noindent {\bf Measuring angular Schmidt spectrum of entangled
states.} The state of the two-photon field produced by parametric
down-conversion (PDC) has the following Schmidt-decomposed form
when the detection system is sensitive only to the OAM-mode index
\cite{jha2011pra}:
\begin{align}
|\psi_{2}\rangle=\sum_{l=-\infty}^{\infty}\sqrt{S_{l}}|l\rangle_{s}|-l\rangle_{i}.
\label{two-photon-state2}
\end{align}
Here $s$ and $i$ stand for signal and idler photons, respectively,
$|l\rangle$ represents a mode with OAM-mode index $l$, and $S_{l}$
is referred to as the angular Schmidt spectrum. The angular
Schmidt spectrum quantifies the dimensionality and the
entanglement of the state in the OAM basis \cite{law2004prl,
torres2003pra, pires2010prl}. There are a variety of techniques
for measuring the angular Schmidt spectrum \cite{mair2001nature,
peeters2007pra, pires2010prl, jha2011pra, giovannini2012njp}. In
the context of spatial entanglement, there has even been a
theoretical proposal \cite{chan2007pra} and its subsequent
experimental implementation \cite{just2013njp} for measuring the
spatial Schmidt spectrum in a single-shot manner using coincidence
detection. However, all the above mentioned work, including the
single-shot work in the spatial domain \cite{chan2007pra,
just2013njp}, are sensitive to noise and require either a very
precise knowledge of the experimental parameters, such as beam
splitting ratio, or a very stable interferometer. In contrast, as
an important experimental application of our technique, we now
report an experimental measurement of the angular Schmidt spectrum
of the PDC photons that not only is a single-shot,
noise-insensitive technique but also does not require coincidence
detection.

As derived in Ref.~\cite{jha2011pra}, the angular coherence
function $W_s(\phi_1,\phi_2)$ corresponding to the individual
signal or idler photon has the following form:
\begin{align}
W_s(\phi_1, \phi_2)\rightarrow
W_s(\Delta\phi)=\frac{1}{2\pi}\sum_{l=-\infty}^{\infty}S_{l}
e^{-il\Delta\phi},\label{OAM-spectrum3}
\end{align}
where $S_l=\sum_{p}\alpha_{l p p}$ is the OAM spectrum of
individual photons. Comparing Equations (\ref{two-photon-state2}) and
(\ref{OAM-spectrum3}), we find that the OAM spectrum of individual
photons is same as the angular Schmidt spectrum of the entangled
state. Therefore, it is clear that one can measure the angular
Schmidt spectrum in a single-shot manner by measuring the
OAM-spectrum of individual photons in a single-shot manner. As
depicted in Fig.~\ref{setup-schematic}(b), entangled photons are
produced by PDC with collinear type-I phase matching. The photons
are collected by a lens arrangement whose collection angle is
larger than the emission cone-angle of the crystal. This ensures
that no part of the produced field is discarded from the
measurement and thus that the true spectrum is measured. This
field is then made incident into the Mach-Zehnder interferometer
of Fig.~\ref{setup-schematic}(b). For a type-I collinear
down-conversion with a 2-mm thick BBO crystal and a $0.85$-mm
beam-waist pump laser, the measured output interferograms and the
corresponding azimuthal intensities for two values of $\delta$
have been shown in Fig.~\ref{spdcpatterns}(a) and
Fig.~\ref{spdcpatterns}(b). Fig.~\ref{spdcpatterns}(c) shows the
normalized measured spectrum as computed using
Equation (\ref{measured-spectrum1}) and the normalized theoretical
spectrum as calculated using the formalism of
Ref.~\cite{torres2003pra}. The near-perfect match without the use
of any fitting parameter shows that we have indeed measured the
true theoretical angular Schmidt spectrum of PDC photons. There is
some mismatch for low-$l$ values, which we attribute to the very
small but finite shot-to-shot noise variation $\Delta I_{\rm
n}(\phi)$. The angular Schmidt number computed as
$K=1/(\sum_{l}\bar{S}^{2}_{l})$ is $K=82.1$, which, to the best of
our knowledge, is the highest-ever reported angular Schmidt number
so far.

\

\noindent {\bf Discussion}

\noindent In conclusion, we have proposed and demonstrated a
single-shot technique  for measuring the angular coherence
function and thereby the OAM spectrum of light fields that can be represented as mixtures of modes with
different OAM values per photon. Our technique involves a Mach-Zehnder interferometer with the
central feature of having an odd and an even number of mirrors in the two interferometric arms, and it is very
robust to noise and does not require precise characterization of
setup parameters, such as beam splitting ratios, degree of
temporal coherence, etc. This technique also does not involve any inherent
post-selection of the field to be measured and thus measures the
true OAM spectrum of a field. As an important application of this
technique, we have reported the measurement of a very
high-dimensional OAM-entangled states in a single shot manner,
without requiring coincidence detection. For such high-dimensional
states, our technique improves the time required for measuring the
OAM spectrum by orders of magnitude. This can have important
implications in terms of improving the signal-to-noise ratio and
reducing the interferometric stability requirements for both
classical and quantum communication protocols that are based on
using OAM of photons. 

The technique presented in this article is for fields that can be
represented as incoherent mixtures of modes carrying different
OAM-mode indices. However, in many important applications, such as
OAM-based multiplexing in communication protocols
\cite{wang2012natphot,bozinovic2013science,yan2014natcom}, one
uses fields that are coherent superpositions of OAM-carrying
modes. For such fields, we believe that the generalizations of the
reconstruction techniques \cite{fienup1987josaa, gaur2015josaa}
used for complex-valued objects could be a possible way of getting
the state information in a single-shot manner. Moreover, in recent years,
finding efficient ways for measuring a partially coherent field is
becoming an important research pursuit \cite{waller2012natphot},
and we believe that, at least in the OAM degree of freedom, the
generalized versions of the existing techniques for coherent
fields in combination with our technique presented in this article
might pave the way towards a full quantum state tomography in a
single-shot or a few-shots manner.

\

\noindent {\bf Methods}

\noindent {\bf Details of the experiment with lab-synthesized
fields.} In this experiment, the $LG_{p=0}^l(\rho, \phi)$ modes
were generated by an SLM using the method by Arrizon {\it et al.}
\cite{arrizon2007josaa}. These modes were made sequentially
incident into the interferometer and the corresponding output
interferograms were imaged using an Andor iXon Ultra EMCCD camera having 512$\times$512 pixels. For each individual $LG_{p=0}^l(\rho, \phi)$ mode the camera was exposed for about 0.4 seconds.
The sequential acquisition was automated to ensure that $\delta$
is the same for all the modes. The azimuthal intensity
$\bar{I}_{\rm out}^{\delta}(\phi)$ plots were obtained by first
precisely positioning a very narrow angular region-of-interest
(ROI) at angle $\phi$ in the interferogram image and then
integrating the intensity within the ROI up to a radius that is
sufficiently large. To reduce pixelation-related noise, the
interferograms were scaled up in size by a factor of four using a
bicubic interpolation method. In order to ensure minimal
shot-to-shot noise variation, the interferometer was covered after
the required alignment with a box and the measurements were
performed only after it had stabilized in terms of ambient
fluctuations.

\

\noindent {\bf Details of the experiment measuring angular
Schmidt spectrum.} As depicted in Fig.~\ref{setup-schematic}(b), a
$405$ nm ultraviolet pump laser with a beam radius 0.85 mm and
having a Gaussian transverse mode profile was made incident on a
2-mm thick beta barium borate (BBO) crystal. The crystal was
phase-matched for collinear type-I PDC. The pump power of $100$ mW
ensured that we were working within the weak down-conversion
limit, in which the probability of producing a four-photon state
is negligibly small compared to that of producing a two-photon
state. The residual pump photons after the crystal were discarded
by means of a dichroic mirror (DM). The down-converted photons
were passed through an interference filter of spectral width $10$
nm centered at $810$ nm and then made incident into the
Mach-Zehnder interferometer of Fig.~\ref{setup-schematic}(b). The
output of the interferometer was recorded using an Andor iXon
Ultra EMCCD camera having 512$\times$512 pixels with the acquisition time of 13 seconds. The
azimuthal intensity $\bar{I}_{\rm out}^{\delta}(\phi)$ plots were
obtained by first precisely positioning a very narrow angular
region-of-interest (ROI) at angle $\phi$ in the interferogram
image and then integrating the intensity within the ROI up to a
radius that is sufficiently large. To reduce pixelation-related
noise, the interferograms are scaled up in size by a factor of
eight using a bicubic interpolation method.

We note that since we are using collinear down-conversion, the
individual signal and idler photons have equal probability of
arriving at a given output port of the interferometer. As a
result, what is recorded by the camera at a given output port is
the sum of the interferograms produced by the signal and idler
fields at that output port. However, since the individual signal
and idler fields have the same OAM spectrum, the azimuthal profile
of the sum interferogram is same as that of the individual
interferograms produced by either the signal or the idler field.
One assumption that we have made here is that the probability of
simultaneous arrivals of the signal and idler photons at the same
EMCCD-camera pixel is negligibly small. This assumption seems
perfectly valid given that the EMCCD camera has 512$\times$512
pixels, and as shown in Fig.~\ref{spdcpatterns}, the output
interferograms occupy more than half of the EMCCD camera pixels.

\

\noindent {\bf Theory of non-symmetric OAM-spectrum measurement.} This section presents our analysis for a
non-symmetric spectrum, that is, when $S_l=S_{-l}$ condition is
not necessarily met. Just as in the case of symmetric spectrum,
let us assume that the measured azimuthal intensity $\bar{I}_{\rm
out}^{\delta}(\phi)$ at the output contains the noise term $I_{\rm
n}^{\delta}(\phi)$ in addition to the signal $I_{\rm out}(\phi)$.
Thus
\begin{align}
&\bar{I}_{\rm out}^{\delta}(\phi)=I_{\rm n}^{\delta}(\phi)+I_{\rm
out}(\phi)\notag\\&=I_{\rm
n}^{\delta}(\phi)+\frac{k_1+k_2}{2\pi}+\gamma \sqrt{k_1k_2} [
W(2\phi)e^{-i\delta}+\rm{c.c.}].
\end{align}
Now, suppose we have two interferogram measured at two different
values of $\delta$, say at $\delta_c$ and $\delta_d$. The
difference $\Delta \bar{I}_{\rm out}(\phi)$ in the intensities of
the two interferogram is then given by
\begin{align}
\Delta &\bar{I}_{\rm out}(\phi)=\bar{I}_{\rm
out}^{\delta_c}(\phi)-\bar{I}_{\rm
out}^{\delta_d}(\phi)\notag\\&=\Delta I_{\rm n}(\phi)+\gamma
\sqrt{k_1k_2} [ W(2\phi)e^{-i\delta_c}+W^*(2\phi)e^{i\delta_c}
\notag\\& \qquad\qquad\qquad\quad-
W(2\phi)e^{-i\delta_d} - W^*(2\phi)e^{i\delta_d}],
\label{output-intensity2}
\end{align}
where $\Delta I_{\rm n}(\phi)=I_{\rm n}^{\delta_c}(\phi) - I_{\rm
n}^{\delta_d}(\phi)$ is the difference in the noise intensities.
Unlike in the case of symmetric spectrum, $\Delta \bar{I}_{\rm
out}(\phi)$ is not proportional to the angular coherence function
$W(2\phi)$. Multiplying each side of Equation (\ref{output-intensity2})
by $e^{i2l\phi}$ and using the angular Wiener-Khintchine relation
$S_l =\int_{-\pi} ^{\pi} W(2\phi) e^{i2l\phi} d(2\phi)$, we obtain
\begin{align}
&\int_{-\pi}^{\pi}\Delta \bar{I}_{\rm
out}(\phi)e^{i2l\phi}d(2\phi)= \int_{-\pi}^{\pi}\Delta I_{\rm
n}(\phi)(\phi)e^{i2l\phi}d(2\phi) 
\notag\\&+ \gamma \sqrt{k_1k_2} [ S_l
e^{-i\delta_c}+S_{-l}e^{i\delta_c}- S_l e^{-i\delta_d} - S_{-l}
e^{i\delta_d}]. \label{output-intensity3}
\end{align}
Now, multiplying each side of Equation (\ref{output-intensity2}) by
$e^{-i2l\phi}$ and using the angular Wiener-Khintchine relation
$S_l =\int_{-\pi} ^{\pi} W(2\phi) e^{i2l\phi} d(2\phi)$, we obtain
\begin{align}
&\int_{-\pi}^{\pi}\Delta \bar{I}_{\rm
out}(\phi)e^{-i2l\phi}d(2\phi)=\int_{-\pi}^{\pi}\Delta I_{\rm
n}(\phi)(\phi)e^{-i2l\phi}d(2\phi)
\notag\\&+\gamma \sqrt{k_1k_2} [ S_{-l}
e^{-i\delta_c}+S_{l}e^{i\delta_c}- S_{-l} e^{-i\delta_d} - S_{l}
e^{i\delta_d}]. \label{output-intensity4}
\end{align}
Adding Equations (\ref{output-intensity3}) and
(\ref{output-intensity4}), we get
\begin{align}
\int_{-\pi} ^{\pi}\Delta & \bar{I}_{\rm
out}(\phi)\cos(2l\phi)d(2\phi)=\int_{-\pi} ^{\pi}\Delta I_{\rm
n}(\phi)\cos(2l\phi)d(2\phi)
\notag\\&+\gamma \sqrt{k_1k_2}
(S_{l}+S_{-l})(\cos\delta_c-\cos\delta_d).
\label{output-intensity5}
\end{align}
Subtracting Equation (\ref{output-intensity4}) from
Equation (\ref{output-intensity3}), we get
\begin{align}
\int_{-\pi}^{\pi}\Delta \bar{I}_{\rm
out}(\phi)&\sin(2l\phi)d(2\phi)=\int_{-\pi}^{\pi}\Delta I_{\rm
n}(\phi)\sin(2l\phi)d(2\phi)
\notag\\&-\gamma \sqrt{k_1k_2}
(S_{l}-S_{-l})(\sin\delta_c-\sin\delta_d).
\label{output-intensity6}
\end{align}
Now the question is how should one define the spectrum so that the
defined spectrum becomes proportional to the true spectrum. Upon
inspection we find that for the non-symmetric case it is not
possible to define the spectrum the way we did it in the case of
symmetric spectrum. Nevertheless, in the special situation in
which $\delta_c + \delta_d = \pi/2$, it is possible to define the
measured spectrum just like we did it in the symmetric case. Let
us consider the situation when $\delta_c=\theta$ and
$\delta_d=\pi/2-\theta$ such that $\delta_c + \delta_d = \pi/2$.
Equations (\ref{output-intensity5}) and (\ref{output-intensity6}) for
this situation can be written as
\begin{align}
\int_{-\pi}^{\pi}\Delta \bar{I}_{\rm
out}(\phi)&\cos(2l\phi)d(2\phi)=\int_{-\pi} ^{\pi}\Delta I_{\rm
n}(\phi)\cos(2l\phi)d(2\phi)
\notag\\&+\gamma \sqrt{k_1k_2}
(S_{l}+S_{-l})(\cos\theta-\sin\theta). \label{output-intensity7}
\end{align}
and
\begin{align}
\int_{-\pi}^{\pi}\Delta \bar{I}_{\rm
out}(\phi)&\sin(2l\phi)d(2\phi)=\int_{-\pi}^{\pi}\Delta I_{\rm
n}(\phi)\sin(2l\phi)d(2\phi)
\notag\\&+\gamma \sqrt{k_1k_2}
(S_{l}-S_{-l})(\cos\theta-\sin\theta). \label{output-intensity8}
\end{align}
Adding Equations (\ref{output-intensity7}) and
(\ref{output-intensity8}), we get
\begin{align}
&\int_{-\pi}^{\pi}\Delta \bar{I}_{\rm
out}(\phi)\left[\cos(2l\phi)+\sin(2l\phi)\right]d(2\phi)\notag\\&=\int_{-\pi}^{\pi}\Delta
I_{\rm
n}(\phi)\left[\cos(2l\phi)+\sin(2l\phi)\right]d(2\phi)
\notag\\& \qquad\qquad+2\gamma
\sqrt{k_1k_2}(\cos\theta-\sin\theta) S_l.
\label{output-intensity9}
\end{align}
So, now if we define the measured spectrum $\bar{S}_l$ to be
\begin{align}
\bar{S}_l&\equiv\int_{-\pi}^{\pi}\Delta \bar{I}_{\rm
out}(\phi)\left[\cos(2l\phi)+\sin(2l\phi)\right]d(2\phi)\notag\\
&=\int_{-\pi/2}^{\pi/2}2\Delta\bar{I}_{\rm
out}(\phi)\left[\cos(2l\phi)+\sin(2l\phi)\right]d\phi\notag\\
&=\int_{-\pi}^{\pi}\Delta\bar{I}_{\rm
out}(\phi)\left[\cos(2l\phi)+\sin(2l\phi)\right]d\phi,
\label{output-intensity10}
\end{align}
we get,
\begin{align}
\bar{S}_l= \int_{-\pi}^{\pi}\Delta I_{\rm
n}(\phi)&\left[\cos(2l\phi)+\sin(2l\phi)\right]d\phi 
\notag\\&+2\gamma
\sqrt{k_1k_2} (\cos\theta-\sin\theta) S_l. \label{spectrum}
\end{align}
In situations in which the noise neither has any explicit
functional dependence on $\delta$ nor has any shot-to-shot
variation, we have $\Delta I_{\rm n}(\phi)=0$. Thus the defined
spectrum $\bar{S}_l$ becomes proportional to the true spectrum
$S_l$. We see that just as in the case of symmetric spectrum, one
does not have to know the exact values of $k_1, k_2, \gamma$ and
$\theta$. The only thing different in this case is that one has to
take the two shots such $\delta_c+\delta_d=\pi/2$. 

\

\noindent {\bf Data availability.} The authors declare that the main data supporting the findings of this study are available within the article. Extra data are available from the corresponding author upon request.

\

\noindent{\bf Acknowledgements}

\noindent We acknowledge financial support through an initiation
grant no. IITK /PHY /20130008 from Indian Institute of Technology
(IIT) Kanpur, India and through the research grant no.
EMR/2015/001931 from the Science and Engineering Research Board
(SERB), Department of Science and Technology, Government of India.

\

\noindent{\bf Author contributions}

\noindent G.K. and A.K.J. proposed the idea, worked out the theory, and performed the
experiments. R.S. contributed towards producing LG beams using the
method by Arriz\'{o}n {\it et al.} \cite{arrizon2007josaa} and
helped with numerical studies. O.S.M.L. and R.W.B. contributed to
the theoretical analysis of the work. A.K.J. supervised the
overall work and all authors contributed in preparing the
manuscript.

\

\noindent{\bf Competing financial interests:} The authors declare no competing financial interests.

\

\noindent{\bf References}


\end{document}